\PassOptionsToPackage{square, numbers, sort}{natbib}
\documentclass[reprint,twocolomn,prr]{revtex4-2}
\usepackage{footmisc}
\usepackage[utf8]{inputenc}
\usepackage[T1]{fontenc}
\usepackage[english]{babel}

\usepackage[fleqn]{amsmath}
\usepackage{mathtools}
\usepackage{amssymb}
\usepackage{xfrac}
\usepackage{physics}

\usepackage{placeins}
\usepackage{graphicx}

\usepackage[percent]{overpic}

\usepackage{booktabs}

\usepackage{tabularx}
\usepackage[graphicx]{realboxes}
\usepackage{array}

\usepackage{xcolor}

\usepackage{ulem}
\usepackage{siunitx}

 \usepackage[utf8]{inputenc}
\usepackage[T1]{fontenc}
 \usepackage{amsmath, amssymb, stmaryrd}
 \usepackage{color, ulem}
 \usepackage{breqn}  % for breaking long equations
 \usepackage{physics}

\begin{document}
 \title{Photonics of topological magnetic textures}
\author{
Vakhtang Jandieri$^{1,2}$, 
Ramaz Khomeriki$^{3}$,
Daniel Erni$^{2}$,
Nicolas~Tsagareli$^{4}$,
Qian Li$^{5}$,
Douglas H. Werner$^{6}$,
and Jamal Berakdar$^{1*}$\\[2ex]
\small
$^{1}$ Institut für Physik, Martin-Luther Universität, Halle-Wittenberg, D-06099 Halle, Germany\\
$^{2}$ General and Theoretical Electrical Engineering (ATE), Faculty of Engineering, University of Duisburg-Essen, and Center for Nanointegration Duisburg-Essen (CENIDE), D-47048 Duisburg, Germany\\
$^{3}$ Physics Department, Tbilisi State University, 3 Chavchavadze, 0128 Tbilisi, Georgia\\
$^{4}$ Department of Electrical and Computer Engineering, Binghamton University, State University of New York, Binghamton, NY 13902, USA\\
$^{5}$ State Key Laboratory of New Ceramics and Fine Processing, School of Materials Science and Engineering, Tsinghua University, Beijing 100084, China\\
$^{6}$ Department of Electrical Engineering, The Pennsylvania State University, University Park, PA 16802, USA\\
*Corresponding author: Jamal.Berakdar@Physik.Uni-Halle.de
}

\begin{abstract}
Topological   textures in magnetically ordered materials are important case studies for fundamental research with promising applications in data science. They can also serve as photonic elements to mold  electromagnetic fields endowing them with features inherent to the spin order, as demonstrated analytically and numerically in this work.  A self-consistent theory is developed for the interaction of spatially structured electromagnetic fields with non-collinear, topologically non-trivial spin textures. A tractable numerical method is designed and implemented for the calculation of the formed magnetic/photonic textures in the entire simulation space.  Numerical illustrations are presented for scattering from  point-like singularities, i.e. Bloch points, in the magnetization vector fields, evidencing  that the geometry and topology of the magnetic order results in photonic fields that embody  orbital angular momentum,  
chirality  as well as magnetoelectric densities.  Features of the scattered fields can serve as a fingerprint for the underlying magnetic texture and its dynamics. The findings point to the potential of topological magnetic textures as a route to molding photonic fields. 
\end{abstract}
\maketitle
\section{Introduction}
Emergence and characteristics  of stable topological textures and quasiparticles such as vortices, skyrmions or hopfions (Hopf solitons) are well studied and documented in the context of field theories, for example in \cite{doi:10.1098/rspa.1961.0018,Faddeev1976,KUZNETSOV198037,PhysRevB.65.100512,PhysRevLett.51.2040,refId0,KOSEVICH1990117,BOGOLUBSKY1988511,Ranada1989,BOGDANOV1994255,Rajaraman1982solitons}. With experimental realizations of these textures in static magnetic and ferroelectric materials as well as in electromagnetic waves \cite{Roeßler2006,doi:10.1126/science.1166767,Yu2010,doi:10.1126/science.1214143,doi:10.1126/science.1240573,Yadav2016,Kent2021,Wiesendanger2016,Shen2024}, much research has been devoted to their possible utilization in information storage, communication, and processing \cite{Dieny2020}. {  In this respect, means are necessary to  controllably exploit, identify, and manipulate their geometric-topological features.} The focus of this work is on  the magneto-optical response as well as on the scattering, and formation of spatially structured electromagnetic fields due to  non-collinear topological textures formed in a magnetization vector field. In particular, we present explicit results and simulations   for point-like magnetic  singularities with vanishing magnetic moments \cite{feldtkeller1965mikromagnetisch,10.1063/1.1656144}. This so-called Bloch-point (BP) singularity can be viewed as a magnetic monopole and  can be surrounded by a hedgehog-type structure or other types of non-collinearities \cite{feldtkeller1965mikromagnetisch,10.1063/1.1656144,kabanov1989bloch,Hertel_2016,PhysRevLett.97.177202,PhysRevB.99.024433}; some examples are detailed below. On the microscopic level the formation of BP is dominated by the exchange interaction and hence the magnetic response should be much faster than the typical GHz magnonic response in homogeneous magnets.\\
BPs have been studied in a wide range of materials and on different length scales, notably  BPs in domains of multilayers and skyrmion lattices or as engineered materials with structuring \cite{kabanov1989bloch,PhysRevB.67.094410,PhysRevLett.97.177202,Blanco-Roldán2015,Hertel_2016,Kanazawa2016,Im2019,PhysRevB.99.024433,PhysRevB.99.024433,Birch2022,PhysRevResearch.2.033006,Rana2023,PhysRevB.108.024414,Zambrano-Rabanal2023,doi:10.1021/acsami.4c01963}. BP singularities are of relevance  also for fundamental   science, for instance  as  emulators of electron motion in a magnetic monopole field \cite{PhysRevB.90.224414}.  For applied science, their distinctive topological \footnote{\label{foot:1} \underline {Topology and geometry of BP:} Considering a  closed surface enclosing  the BP at its center,  the direction of unit magnetization vector field ($\mathbf{m}_0$ in Eq.\ref{eq::BP})   covers  an integer
number of times, a whole solid angle.  Expressed in spherical coordinates \cite{10.1063/1.1656144}, this vector field is parameterized by a polar component expressible as $p\theta$, and  an azimuthal component written as  $q\phi + \gamma$. Since $\mathbf{m}_0$ is single valued, $q$ and $p$ are integer number and the topological nature of the BP is then set by the quantity  $Q=pq$, called the topological charge \cite{Rajaraman1982solitons}.  $\gamma$ determines the geometry: It is an azimuthal tilt with regard  to the radial direction, causing  the BP to twist around $z$, as discussed  in the text and shown in Fig.(\ref{fig::3}). Such as behavior  and how the system goes from one $q,p$ state to another is  illustrated  for example in \cite{Jia_2019}  for  vortices and skyrmions using micromagnetic  simulations. The discussions and numerical calculations presented  in this work assumes  a specific $\gamma$. The actual value of $\gamma$ for a given material depends on terms in the magneto-static energy, such as
the dipolar energy  \cite{10.1063/1.1656144}, that  are not in the focus of this work.}
nanometer scale,  as well as energy scale \footnote{\label{foot:2}\underline{ Energy scale}: In the region around the BP the dominant energy contribution stems
from  exchange energy $\int d^3\mathbf{r} \frac{J}{2a} (\nabla \mathbf{S})$, where $J$ is the exchange interaction, $a$ is the lattice constant and $\hbar S$ is the molecular spin  which is proportional to  the magnetization $\mathbf{M}$ appearing in (\ref{eq:2}). The energy scale is typically an order of magnitude higher than that due to dipolar interactions which govern magneto-static excitations (spin waves) in uniform magnetic media.} features make them a compelling option for data storage and processing \cite{kabanov1989bloch,PhysRevB.67.094410,PhysRevLett.97.177202,Blanco-Roldán2015,Hertel_2016,Kanazawa2016,Im2019,PhysRevB.99.024433,PhysRevB.99.024433,Birch2022,PhysRevResearch.2.033006,Rana2023,PhysRevB.108.024414,Zambrano-Rabanal2023,doi:10.1021/acsami.4c01963}. 
To characterize and drive controllably  Bloch points is  however a challenge, mainly due to their vanishing magnetization. Clearly, footprints of their topological characteristics are left  in their surrounding long-range magnetic order.  Hence, a probe which couples directly to their intrinsic non-collinear structure would be highly desirable. Also the question of a possible use of BPs or similar textures  as functional photonic elements needs to be clarified.

In this work we study the scattering and formation  of structured electromagnetic (EM) waves from BPs, and illustrate the polarization distribution  and the  spatial  structure of the fields, as reflected for example by  the chirality density and/or the orbital angular momentum density. While propagating EM waves are bound in their spatial resolution to the diffraction limit, there is no such limit on the polarization texturing \cite{Zhan:09,Rosales-Guzman_2018,Grunwald01012020,10.3389/fphy.2021.688284,Werner,Kozawa2025250202,Vogliardi2025}. In fact, the EM fields used here can also be singular beams with the singularity occurring on the optical axis. In this way a spatial resolution on the nanometer scale of the magnetic singularity can be achieved. In recent years clear experimental and theoretical evidence has been accumulated on how to use singular of spatially structured fields to access new information on matter \cite{PhysRevLett.134.156701,PhysRevLett.128.157205,Rouxel2022,doi:10.1021/acs.chemrev.2c00115,Bose_2014,DeNinno2020,LinNieYanLiangLinZhaoJia+2019+2177+2188,PhysRevA.103.013501,PhysRevLett.128.077401,PhysRevLett.132.026902,Waetzel2018,https://doi.org/10.1002/qute.201800096,RevModPhys.94.035003}, but so far a closed theory describing light-matter action-back-action, particularity in the context of magnetically ordered systems remains elusive. Such a theory would bridge the gap between topological photonics \cite{RevModPhys.91.015006} and  magneto photonics \cite{Wang2009} to topological magnetic textures. \\

An important issue to clarify is how the singular magnetic texture responds to external EM fields and how the  fields are modified by the magnetic texture, when the frequencies of these EM fields are in the range of magnetic excitations.
% {  X-rays have been extensively used to study magnetic topological textures, such as skyrmions and domain walls. However, their application to Bloch points remains limited due to the latter’s nanoscale size and inherently three-dimensional nature. In this paper, we instead propose to monitor the nontrivial electromagnetic topological response in the frequency range corresponding to Bloch point resonances.} 
Therefore, we will set up a coupled, self-consistent magnetic-photonic modeling scheme for  the case involving a large photon number (classical EM fields) and where the length  (energy) scales of relevance are much larger (smaller) than the atomistic ones. Under these conditions it is reasonable to work within  classical field theories and incorporate the atomistic electronic information, as is customary with dielectric/permeability  tensors. 

To be explicit we focus here on BPs as a paradigm for highly non-collinear systems. Our theoretical formulation is however 
applicable to other types of nonlinearities    such as vortices \cite{fanciulli2025magnetic,luttmann2025optical}, skyrmions, spin ice \cite{10.1063/5.0274799,10.1063/5.0229120}, gyroids \cite{doi:10.1021/acsami.4c02366}, or  external magnetic fields imparted by other structured waves \cite{Geerits2023,PhysRevB.107.134403,Henderson2023,Sarenac2024}.
The specific system is manifested in the appropriate  magnetic free energy density and the corresponding
boundary conditions. Generally, the magnetic free energy density  contains both the external magnetic field $\boldsymbol{H}_{\text{ext}}(\boldsymbol{r},t)$ of the incoming EM waves and contributions from intrinsic magnetic interactions $\boldsymbol{H}(\boldsymbol{r},t)$ such as exchange  coupling. The dynamics of the local magnetization $\mathbf{M}(\boldsymbol{r},t)$, described by the Landau-Lifshits-Gilbert equation implicitly updates for the magnetic field  %delivers the updated magnetic field 
%$\mathbf{H}(\mathbf{r}',t')$ 
at time $t'$ and space $\boldsymbol{r}'$ to be used as an input value for Maxwell's equations to infer the electric flux density $\mathbf{D}(\boldsymbol{r}',t')$. The successive applications of this procedure accounts for multiple magnetic/electromagnetic scattering events. In the linear response regime, we analytically derive 
%to $H$, 
 the frequency and wave-vector-dependent permeability tensor that can be used for the low-energy  dynamics of the magnetic textures. A time-harmonic dependence of $e^{j\omega t}$ is assumed and omitted throughout the manuscript.
 
 % Although we discuss here numerical simulations for BPs, the  formalism is more general and can be applied to other types of topological magnetic objects such as vortices \cite{fanciulli2025magnetic,luttmann2025optical}, skyrmions, spin ice \cite{10.1063/5.0274799,10.1063/5.0229120}, or  external magnetic fields imparted by other structured waves \cite{Geerits2023,PhysRevB.107.134403,Henderson2023,Sarenac2024}.   

% ---- JB should add here the reference to Ruchon's work and also mention the experiment on vortex dynamics in structured fields, also mention use of structured fields as very localized magnetic fields (Nature paper) ----

%The work is organized as follows.

\section{Formal considerations}
For non-collinear magnets, we  seek a self-consistent spherical rigorous-coupled wave analysis for EM wave scattering in three-dimensions (3D). For instance, the theory should account for the  inhomogeneities and {  dynamical response of the magnetic order spherically surrounding the BP (see the inset in Fig. \ref{fig::1})}. In the linear response regime, we wish to analytically obtain a  magnetic permeability matrix and uncover the variation  of it components along the radial direction and along the polar angle. Hereby, we  assume azimuthally homogeneous BPs. This is accomplished below where, technically, the method reduces the problem to the  solution of a set of  first-order coupled differential equations. Moreover, the field representations are then found using the eigenvectors and eigenvalues of the matrix generated by the Fourier components of the constant coefficients of the differential equations. Numerically, the approach uses the recursive algorithm based on  matrix multiplication, which allows us to calculate the scattering characteristics for the BP efficiently. The method can deal numerically with various types of impinging electromagnetic waves and BPs such as  hedgehog and twisted. For brevity only selected   results are presented  showing that the distinct magnetic order around the BP leads to an involved structure of the formed  EM fields 
and the generation of optical chirality density and orbital angular momentum  in a close vicinity to the BP. 
Also, we find a dominant  azimuthal component of the Poynting vector   close to the surface of the BP. Only the components of the reflected electric and magnetic fields that appear solely by an interaction of the incident waves with the BP show a sensitive behavior while transforming from the hedgehog to the twisted model.

%A time-harmonic dependence of $e^{j\omega t}$ is assumed
%and omitted throughout the manuscript.
%--- currently the topic of spatio-temporal photonic materials is pretty much discussed. Here, we have also this case if we keep the two times, the one from the field and the one from the LLG. There is no need to calculations but it would give the work much more value if we start from the most general case, derive the eqaution and then go the linearized form and then to frequency space ...

%\subsection{Temporal dynamics of interaction of the Bloch points and EM waves}
The formulation below is developed for a magnetic insulator, meaning that no free currents/charges are present, and assumes that the dielectric response  (and hence the relative dielectric permittivity  $\epsilon_r$)
is nearly constant and homogeneous  in the frequency regime of interest here. These restrictions are not fundamental  but allow the magnetic/photonic effects to be singled out. The coupled field equations of motion for magneto-static/electromagnetic dynamics are given by
\begin{equation}
\begin{aligned}
\frac{\partial \boldsymbol{M}(\mathbf{r},t)}{\partial t} = -g\mu_0\left[\boldsymbol{M}(\mathbf{r},t) \times \boldsymbol{H}_{\text{eff}}(\mathbf{r},t)\right], \\
 \nabla\times \left[\boldsymbol{H} (\mathbf{r},t)+ \boldsymbol{H}_{\text{ext}}(\mathbf{r},t)\right] = \epsilon_0 \epsilon_r \frac{\partial \boldsymbol{E}(\mathbf{r},t)}{\partial t}, \\
 \nabla\times \boldsymbol{E} (\mathbf{r},t) = -\mu_0  \frac{\partial} {\partial t}\left[\boldsymbol{H}(\mathbf{r},t)+\boldsymbol{H}_{\text{ext}}(\mathbf{r},t)+\boldsymbol{M}(\mathbf{r},t)\right]
\end{aligned}
\label{eq:full}\end{equation}
where $g$ is the gyromagnetic ratio, $\boldsymbol{H}$ and $\boldsymbol{H}_{\text{ext}}$ are the demagnetization and external magnetic fields, respectively, $\boldsymbol{M}$ is the magnetization vector field, $\epsilon_0$ and $\mu_0$ are dielectric permittivity and magnetic permeability of free space. The electric polarization induced by BP spin-noncollinearities (e.g., as in \cite{PhysRevLett.127.127601}) is negligible. {  The external magnetic field $\boldsymbol{H}_{\text{ext}}$ is related to the incident field and is defined by the boundary condition on a surface of the sphere enclosing the the magnetic singularity (see the inset of Fig. \ref{fig::1}).} 
%In Sec. III we are considering an elementary dipole as a source for
%the lowest-order incident spherical waves. More details are given below in Section II.
%This set of equations describes a time dynamics of an interaction of the BP with the EM waves. 

The effective magnetic field $\boldsymbol{H}_{\text{eff}}$ derives from the functional derivative of the free energy $F(\boldsymbol{M})$  with respect to $\boldsymbol{M}$. {  For BP an appropriate model  $F(\boldsymbol{M})$  may be represented in the form \cite{elias2011magnetization},
\footnote{Directly at the BP where the magnetization vanishes (cf. Fig.\ref{fig::1})  atomistic models are needed. The theory  presented here does not include this particular point. A reasonable length beyond which our model is expected to hold is set by the 
exchange length $\sqrt{J/(M_s^2 a)}$, the value of this length is $\approx  6$ nm for Permalloy. Since our focus is not on describing BPs as such but on the behavior of the EM fields scattering from BPs, and the fields do not scatter when $M_s=0$, this issue is less critical for this study.}
\begin{equation}
\begin{aligned}
F(\boldsymbol{M}) &= \int_V d^3\mathbf{r} \left[
\frac{A_s}{2} (\nabla \boldsymbol{M})^2 + \gamma_1M^2 + \gamma_2M^4 \right. \\
& \left. - \frac{\mu_0}{2} \boldsymbol{M} \cdot (\boldsymbol{H} + \boldsymbol{H}_{\text{ext}}) \right].
\end{aligned}
\label{eq:2}
\end{equation}
The effective field is    
\begin{equation}
\begin{aligned}
%\boldsymbol{H}_{\text{eff}} &= \boldsymbol{H} + \boldsymbol{H}_{\text{ext}} + \left(
%\frac{\partial^2}{\partial r^2} + \frac{2}{r} \frac{\partial}{\partial r} - \frac{2}{r^2}
%\right) \boldsymbol{m}(\theta)M(r) \\
%&+ 2\nu \boldsymbol{m}(\theta)\left( M - \beta M^3 \right)
\boldsymbol{H}_{\text{eff}} &=-\frac{\delta F}{\delta \boldsymbol{M}}=\boldsymbol{H}_{\text{ext}} + \boldsymbol{H} + \left(
\frac{\partial^2}{\partial r^2} + \frac{2}{r} \frac{\partial}{\partial r} - \frac{2}{r^2}
\right) \boldsymbol{M} \\
&+ 2\nu \left( \boldsymbol{M} - \beta\boldsymbol{M}^3 \right)
\end{aligned}
\label{eq:3}
\end{equation}
where $A_s$ is related to the exchange interaction and $\gamma_1,\gamma_2$ are the Landau parameters. We introduce the definitions 
$\nu=|\gamma_1|/\mu_0$, $\beta=-2\gamma_2/\gamma_1$, and  use $r$ as the distance from the global origin, i.e. the center of the BP, to the observation point.} 

We seek a material-independent  formulation but  as a typical example we consider here, the helimagnet FeGe for which the exchange stiffness $A=A_s/2$ is $6.1\; 10^{-13}$ cm$^2$;  the saturation magnetization  is $M_s=383$  emu/cm$^3$, and the Curie temperature 
is $278.7$ K \cite{Lang2023}. These parameters  vary strongly if we consider   a different setting such as spin ice \cite{10.1063/5.0274799,10.1063/5.0229120}.

%\footnote{We seek a general formulation. Hence, we do not discuss in details specific materials. As a typical example we mention however  the helimagnet FeGe for which the exchange stiffness $A=A_s/2$ is $6.1\; 10^{-13}$ cm$^2$;  the saturation magnetization  is $M_s=383$  emu/cm$^3$, and the Curie temperature 
%is $278.7$ K \cite{Lang2023}. These parameters  vary strongly if we consider   a different setting such as spin ice \cite{10.1063/5.0274799,10.1063/5.0229120}.}.

\subsection{Local magnetic susceptibility}
The non-linear coupled equations (\ref{eq:full}) account for a multitude of phenomena such as large-angle magnetic procession/reversal (magnetic moment quenching is not included in these equations), trapping of electromagnetic waves, and frequency up- and down-conversion. In the linear regime, information on the sample in its field-free state can be obtained. The task here is   to  derive
 explicitly  an expression for the magnetic permeability matrix of the BP, which as  we expect from the above discussion should  be frequency dependent and non-local.   To this end, we  linearize $\boldsymbol{M}$ with respect to small deviations caused by the 
 {  external field. We consider a time-independent ground state as the free energy \eqref{eq:2} minimum $\boldsymbol{H}_{\text{eff}}=0$ \cite{elias2011magnetization} 
 %, i.e. getting $\boldsymbol{H}_{\text{eff}}=0$ and supplementing by magnetostatic equations (last two equations of \eqref{eq:full} without external field and time derivatives). The corresponding ground states are derived in \cite{elias2011magnetization} 
 and examine small deviations of these ground state solutions under a Landau–Lifshitz–Gilbert formulation}. We substitute the expression for the effective magnetic field $\boldsymbol{H}_{\text{eff}}$ into the Landau–Lifshitz–Gilbert equation and write  \footnote{Our formulation based on Eqs.(\ref{eq:full}, \ref{eq:2}) captures the physics of BPs. It does not include the magnetic material well beyond BP. This part enters our theory/simulations through a finite but constant relative permeability.  Field matching is performed between the BP region and the surrounding rendering the EM fields well-defined in the whole space. Note, typically  the response of the exchange-dominated BP is at a different (higher) frequency than the magneto-statically-coupled region with a response in the GHz. Mathematically this means that within our formulation and boundary conditions, we cannot realize 
 BP  as localized solutions in a  vector  field
 which is spatially homogeneous, meaning collinear at infinity.}
\begin{equation}
    \boldsymbol{M}(\boldsymbol{r},t) = \boldsymbol{M}_0 + \delta \boldsymbol{M} =   \boldsymbol{m}_0(\boldsymbol{r},t) M_0(r) + \boldsymbol{m}_1(\boldsymbol{r},t)M_0(r). \label{eq::BP}\end{equation}
  Here, $\boldsymbol{{M}}_0(\boldsymbol{r},t)$ is the {  ground state} spin configuration which is described by local magnetization  magnitude $M_0(r)$ and a unit vector field $\boldsymbol{m}_0(\boldsymbol{r},t)$. Small deviations from the initial configuration are captured by  the vector field $\delta \boldsymbol{M}(\boldsymbol{r},t) = \boldsymbol{m}_1(\boldsymbol{r},t)M_0(r)$. 
%(note, the amplitude (longitudinal) magnetic excitations are energetically higher than the local processional motion  described by $\boldsymbol{m}_1(\boldsymbol{r},t)$).  
To first-order in $\boldsymbol{m}_1$  we find
\begin{equation}
\begin{aligned}
\frac{\partial \boldsymbol{m_1}}{\partial t}
=-g\mu_0M_s[\boldsymbol{m_0} \times \boldsymbol{H}_{\text{eff}}].
\end{aligned}
\label{eq:4}\end{equation}
In spherical coordinates with the center being at the BP singularity we write
\begin{equation}
\begin{aligned}
m_r^0 &= \sin^2 \theta \cos \gamma + \cos^2 \theta ,\\
m_\theta^0 &= \cos \theta \sin \theta (\cos \gamma - 1) ,\\
m_\varphi^0 &= \sin \gamma \sin \theta.
\end{aligned}
\label{eq:5}
\end{equation}
%Assuming that $\boldsymbol{M}(r,\theta) = \boldsymbol{m}(\theta)M(r)$ is independent of an azimuthal angle $\varphi$ and
Here $\gamma$ is a pseudo azimuthal  angle which determines the geometry of the spin configuration without changing the topology. For example, for one magnetic singularity at the origin, one transforms from the hedgehog configuration for  $\gamma=0$ to a twisted one for $\gamma=60^{\rm o}$, 
where $\theta$ is the polar angle between the radial line and the polar $z$-axis. $M_0$ is space dependent but still we can normalize to the maximum value of the saturation magnetization $M_s$, as done in (\ref{eq:4}), implying  $\beta M_s^2 =1$. We note that no $\varphi$-dependence is present and $\boldsymbol{r} = (\theta,r)$.    

For a time-harmonic,  polarization structured (vector) EM incident beam we start from (\ref{eq:4}) and seek the elements of the magnetic permeability matrix $\mu_{ij}(\theta,r) = \delta_{ij}+M_0(r)\chi_{ij}(\theta,r)$ (where $\chi_{i,j}(\theta,r)$ is a magnetic susceptibility  $\boldsymbol{m_1} = \chi_{i,j}(\boldsymbol{H} + \boldsymbol{H}_{\text{eff}})$). It is possible to derive the local element  $\chi_{ij}(\theta,r)$ explicitly. The details involve relatively lengthy expressions which are given in Appendix A. Note that the magnetic permeability tensor is  Hermitian.  As detailed in Appendix A,
a key signature of the spin non-collinearity is the appearance of multiple local resonances $\omega _{res}$ determined by the condition 
\begin{eqnarray}
  \omega _{res}&=&  \pm  \mathrm{g} \mu_0 M_s\left[\hat{O}M_0(r)\right],\label{eq::reson}\\
    \hat{O} &=& \left(\frac{\partial^2}{\partial r^2} + \frac{2}{r} \frac{\partial}{\partial r} - \frac{2}{r^2}\right) + 2\nu \left(1 - M_0^2\right).\nonumber
\end{eqnarray}
Importantly, this fact can be exploited to sense for BPs in the frequency space without being hindered by the limited spatial resolution of optical probes.  

\subsection{Scattering of electromagnetic fields: Spherical coupled-wave analysis}
\begin{figure}[b]
\centering \includegraphics[width=\linewidth]{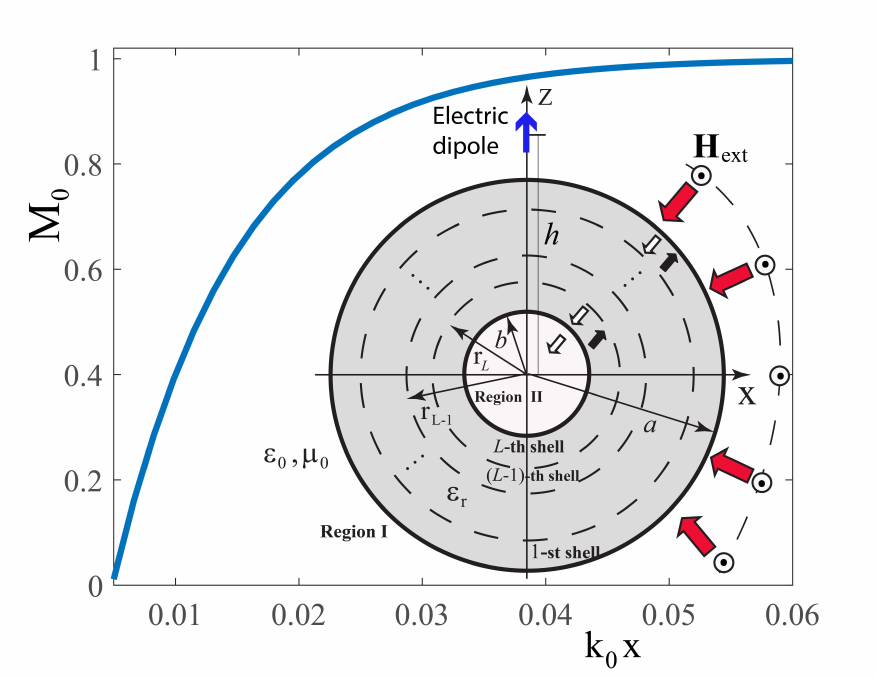}
\caption{Curve showing the normalized magnetization $M_0$ as a function of $k_0x$, where $k_0$ is the free-space wavenumber. {  Inset shows a schematic cross-sectional view of the BP, where the magnetic order is spherically surrounding the BP. The outer-radius of the sphere is $a$ and the inner-radius is denoted as $b$. The anisotropic region (see Appendix A) is marked by the gray color.  It is divided into thin shell layers, each of which is characterized by the local magnetic permeability matrix $\mu_{ij}(\theta,r_l)$. Outside the anisotropic structure the medium is assumed to be free space.} The relative dielectric permittivity of the spherical BP is $\epsilon_r$. The wavefront of the incident field is also depicted by thick red arrows. The direction of the associated magnetic field $\mathbf{H}_{ext}$ is indicated by dotted circles. Such a field can be realized by  a radially directed elementary electric dipole which is  marked by a blue arrow and is located on and parallel to the $z$-axis at a distance $h$ from the global origin (i.e., the center of the BP).  }
\label{fig::1}
\end{figure}

To calculate the scattered fields from localized  magnetic non-collinearities, we formulate a spherical rigorous coupled-wave analysis by dividing the sphere around the magnetic singularity  into onion-type,  thin shell layers, where the $l$-th thin layer $(l=1 \cdot \cdot \cdot L)$ is characterized by a  magnetic permeability matrix $\mu_{ij}(\theta, r_l)$ (c.f. an inset in Fig. \ref{fig::1}). Here, $ r_l$ is the radial distance of the  $l$-th thin shell layer and all material parameters are assumed to be constant within each layer. It is important to mention that each shell layer resonates at a different frequency defined by (\ref{eq::reson}). 
Hence, the magneto-optical behavior under  irradiation with a  broadband vector beam, such as suggested in \cite{doi:10.1021/acsphotonics.1c01693}, will be qualitatively different from a homogeneous beam. For the numerical demonstration  we choose the excitation frequency $\omega$ close to the resonance frequency of the outermost shell layer, i.e. $\omega_{res}(r=a)$. This means that the elements of the magnetic permeability matrix  attain maximum values in the outermost layer and then decrease within the BP when approaching its global origin (detailed derivation is given in Appendix A).

Having  the magnetic permeability matrix  rigorously derived, we then substitute it into the Maxwell's equations written in  spherical coordinates and redefine  the  components of the electric $E_r^{\alpha}$ and magnetic $H_r^{\alpha}$ fields  as $$ S^{\alpha}_{\theta} = r E^{\alpha}_{\theta}, \, S^{\alpha}_{\varphi} = r \sin \theta E^{\alpha}_{\varphi}, $$  $$  U^{\alpha}_{\theta} = \sqrt{{\mu_0}/{\epsilon_0}}\; r  H^{\alpha}_{\theta},U^{\alpha}_{\varphi} = \sqrt{{\mu_0}/{\epsilon_0}}\;  r \sin \theta H^{\alpha}_{\varphi},$$  where ${\alpha}=inc,r$ represents the incident or reflected fields. {  By expanding  the fields into a one-dimensional (1-D)  Fourier series with respect to a parameter $\kappa = \cos \theta$ ($\theta = [0:\pi]$) \cite{jarem1,jarem2}:
\begin{equation}
\Phi(r, \theta) = \sum_{i=-N}^{N} \phi_i(r) e^{ji\pi \kappa}, \quad
\Phi = \{ S_{\theta}, S_{\varphi}, U_{\theta}, U_{\varphi} \}
\label{eq:7}
\end{equation}
a set of  first-order, coupled differential equations for the radially varying field amplitudes $\phi_i(r) = [{s}_{\theta}^i, {s}_{\varphi}^i, {u}_{\theta}^i,{u}_{\varphi}^i]$ is obtained}:   
\begin{equation}
\frac{d}{d\eta}
\begin{bmatrix}
{s}_{\theta}^i \\
{s}_{\varphi}^i \\
{u}_{\theta}^i \\
{u}_{\varphi}^i
\end{bmatrix}
= j \begin{bmatrix}
    \boldsymbol{\Psi}
\end{bmatrix} =
j 
\begin{bmatrix}
\Psi_{11} & \Psi_{12} & \Psi_{13} & \Psi_{14} \\
\Psi_{21} & \Psi_{22} & \Psi_{23} & \Psi_{24} \\
\Psi_{31} & \Psi_{32} & \Psi_{33} & \Psi_{34} \\
\Psi_{41} & \Psi_{42} & \Psi_{43} & \Psi_{44}
\end{bmatrix}
\begin{bmatrix}
{s}_{\theta}^i \\
{s}_{\varphi}^i \\
{u}_{\theta}^i \\
{u}_{\varphi}^i
\end{bmatrix}.
\label{eq:8}
\end{equation}
Here $\eta=k_0r$, ${k}_{0}$ is the free-space wavenumber,  $N$ is a truncation number of the Fourier series that will be varied to check convergence, $\boldsymbol{\Psi}$ is a square matrix generated by the Fourier coefficients of $\Psi_{qq'}$ with the $(n,m)$ entries equal to $\Psi_{qq'}^{n-m}$. If a truncation number for each field component associated with the Fourier series is $N$, ($-N\cdot \cdot \cdot 0 \cdot \cdot \cdot N$), then the size of the matrix $\boldsymbol{\Psi}$ becomes $4(2N+1)$$\times$$4(2N+1)$. The implicit expressions of block matrices $\Psi_{qq'}$ are given in Appendix B. 

Using (\ref{eq:8}), the fields for each anisotropic layer are  expressed through the eigenvectors and eigenvalues of the matrix $\boldsymbol{\Psi}$.  Hence, the tangential components of the electric and magnetic fields in the $l$-th shell layer characterized by a normalized radial distance $\eta_l = k_0 r_l$, namely $({\boldsymbol{S}_\theta}^{(l)}({\kappa,\eta_l}), {\boldsymbol{S}_\varphi}^{(l)}({\kappa,\eta_l}))$ and $({\boldsymbol{U}_\theta}^{(l)}({\kappa,\eta_l}), {\boldsymbol{U}_\varphi}^{(l)}({\kappa,\eta_l}))$, can be written in the following form: 
{\setlength{\mathindent}{8pt}
\begin{align}
{\textbf{S}}^{(l)}({\kappa}, \eta_l)
 &= \sum_{i=-N}^{N}
    \Bigg(
        \sum_{p=1}^{4(2N+1)}
        P_{ip,S}^{(l)}\, e^{j\eta_l |\xi^l_p|}\, C_p^{(l)}
    \Bigg)
    e^{ji\pi\kappa} \nonumber\\[4pt]
 &\quad =
    \sum_{i=-N}^{N}
    \Big(
        \overline{\overline{\boldsymbol{P}}}_S^{(l)} \cdot
        \boldsymbol{\Upsilon} \cdot
        \boldsymbol{C}^{(l)}
    \Big)
    e^{ji\pi\kappa}.
\label{eq::9}
\end{align}

\begin{align}
{\textbf{U}}^{(l)}({\kappa}, \eta_l)
 &= \sum_{i=-N}^{N}
    \Bigg(
        \sum_{p=1}^{4(2N+1)}
        P_{ip,U}^{(l)}\, e^{j\eta_l |\xi^l_p|}\, C_p^{(l)}
    \Bigg)
    e^{ji\pi\kappa} \nonumber\\[4pt]
 &\quad =
    \sum_{i=-N}^{N}
    \Big(
        \overline{\overline{\boldsymbol{P}}}_U^{(l)} \cdot
        \boldsymbol{\Upsilon} \cdot
        \boldsymbol{C}^{(l)}
    \Big)
    e^{ji\pi\kappa}.
\label{eq::10}
\end{align}
}

where we introduced the shorthand notation 
\begin{align}
\bm{\textbf{S}}^{(l)}(\bm{\kappa,\eta_l}) &= [\bm{\boldsymbol{S}_\theta}^{(l)}(\bm{\kappa,\eta_l}),\bm{\boldsymbol{S}_\varphi}^{(l)}\bm{(\kappa,\eta_l})]^T, \\
\bm{\textbf{U}}^{(l)}(\bm{\kappa,\eta_l}) &= [\bm{\boldsymbol{U}_\theta}^{(l)}(\bm{\kappa,\eta_l}),\bm{\boldsymbol{U}_\varphi}^{(l)}(\bm{\kappa,\eta_l})]^T     .
\end{align}
Here, the matrices $\overline{\overline{\bm{\boldsymbol{P}}}}_S^{(l)}$ and $\overline{\overline{\bm{\boldsymbol{P}}}}_U^{(l)}$  having a size $2(2N+1)$$\times$$4(2N+1)$ are constructed from the eigenvectors of the matrix $\boldsymbol{\Psi}^l$ in the $l$-th shell layer.  $\{\xi^l_p\}$ determine the elements of the diagonal matrix $\boldsymbol{\Upsilon}$
 and are set by the respective eigenvalues of $\boldsymbol{\Psi}^l$. 

A relation between the unknown coefficients ${\boldsymbol{C}^{(l)}}$ and  ${\boldsymbol{C}^{(l-1)}}$ in the $l$-th and $(l-1)$-th shell layers, respectively, is determined by the boundary conditions for the incoming and outgoing waves on the  $l$-th layer's boundary \cite{jarem2, jandieri2}:
\begin{equation}
\begin{aligned}
{\boldsymbol{C}}^{(l)} = [{\boldsymbol{P}}^{(l)}]^{-1} \cdot {\boldsymbol{P}}^{(l-1)} \cdot {\boldsymbol{T}}^{(l-1)} \cdot {\boldsymbol{C}}^{(l-1)}  \equiv \mathbf{F}^l \cdot {\boldsymbol{C}}^{(l-1)}
\end{aligned}
\label{eq::13}
\end{equation}
Here ${\boldsymbol{P}}^{(l)} = [\overline{\overline{\bm{\boldsymbol{P}}}}_S^{(l)}, \overline{\overline{\bm{\boldsymbol{P}}}}_U^{(l)}]^T$ is a square matrix (its size is $4(2N+1)$$\times$$4(2N+1)$) whose columns are the eigenvectors of the matrix $\boldsymbol{\Psi}^{l}$ in the $l$-th shell layer, ${\boldsymbol{T}}^{(l-1)}$ is related to the thickness of the $(l-1)$-th thin shell and represents the traveling wave behavior of the space-harmonics along the radial direction. Since the relationship between the coefficients for the adjacent layers inside the sphere have been determined, the relation between $\bm{\boldsymbol{C}}^{(1)}$ in the outermost $1$-st layer inside the sphere (the nearest one to the region of an incident wave) and  $\bm{\boldsymbol{C}}^{(L)}$ in the innermost $L$-th layer (the closest one to the global origin) can be calculated using a recursive algorithm based on the matrix multiplication: $\bm{\boldsymbol{C}}^{(L)} = \mathbf{F} \cdot \bm{\boldsymbol{C}}^{(1)}$. The matrix $\mathbf{F}$, which is a product of matrices ${\mathbf{F}^l} (l = 1 \cdot \cdot \cdot L)$, rigorously accounts for  multiple interaction of the space-harmonics between the shell layers, as well as their material and geometrical parameters through the eigenvectors and eigenvalues of the $\boldsymbol{\Psi}^l$ matrix.   

The electromagnetic fields in the  space $r > a$, i.e., Region I in the inset of Fig. \ref{fig::1}, are expressed in terms of transverse to $r$ electric ($TE_r$) and transverse to $r$ magnetic ($TM_r$) waves and may be written using Schelkunoff-Bessel and Hankel functions in the following form:
\begin{subequations}
\begin{align}
S^{(\text{I})}_{\theta}(r, \kappa) &=  \sum_{i=-N}^{N} (\mathbf{K} \cdot \boldsymbol{c})\, e^{j i \pi \kappa}
 + \sum_{i=-N}^{N} \bm{S}^{\text{inc}}_{\theta,i}\, e^{j i \pi \kappa} \label{eq:S_theta} \\
S^{(\text{I})}_{\varphi}(r, \kappa) &= - \sum_{i=-N}^{N} (\tilde{\mathbf{K}} \cdot \boldsymbol{d})\, e^{j i \pi \kappa}
 + \sum_{i=-N}^{N} \bm{S}^{\text{inc}}_{\varphi,i}\, e^{j i \pi \kappa} \label{eq:S_phi} \\
 U^{(\text{I})}_{\theta}(r, \kappa) &= \sum_{i=-N}^{N} (\mathbf{K} \cdot \boldsymbol{d})\, e^{j i \pi \kappa}
+ \sum_{i=-N}^{N} U^{\text{inc}}_{\theta,i}\, e^{j i \pi \kappa} \label{eq:U_theta} \\
U^{(\text{I})}_{\varphi}(r, \kappa) &= \sum_{i=-N}^{N} (\tilde{\mathbf{K}} \cdot \boldsymbol{c})\, e^{j i \pi \kappa}
 + \sum_{i=-N}^{N} U^{\text{inc}}_{\varphi,i}\, e^{j i \pi \kappa} \label{eq:U_phi}
\end{align}
\end{subequations}
with
\begin{align}
&\mathbf{K} = \{K_{in}\} = j \mathrm{g}_{in} \hat{h}^{(2)'}_n (\eta) ,\label{eq:15}\\
&\tilde{\mathbf{K}} = \{ \tilde{\mathrm{K}}_{in} \}= \tilde{\mathrm{g}}_{in} \hat{h}^{(2)}_n (\eta)  ,\label{eq:16}\\
&\mathrm{g}_{in} = \frac{1}{2} \int_{-1}^{1} d\kappa \sqrt{1 - \kappa^2} \frac{dP^0_n(\kappa)}{d\kappa} e^{-j i \pi \kappa},  \label{eq:17}\\
&\tilde{\mathrm{g}}_{in} = \frac{1}{2} \int_{-1}^{1} d\kappa (1 - \kappa^2) \frac{dP^0_n(\kappa)}{d\kappa} e^{-j i \pi \kappa}.  \label{eq:18}
\end{align}
 Here $P^0_n(\kappa)$ are the associated Legendre polynomials, $\hat{h}^{(2)}_n(\eta)$ are the Schelkunoff Hankel functions of the second kind  satisfying the radiation condition and the prime denotes  differentiation with respect to the argument.  $\boldsymbol{c}$ and $\boldsymbol{d}$ are the scattering amplitudes that should be determined by matching the boundary conditions at $r=a$. The first terms on the right side in (\ref{eq:S_theta})-(\ref{eq:U_phi}) represent the reflected waves, whereas the second terms are the incident waves. The $i$-th Fourier component of the radial electric and magnetic fields can be written as 
\begin{equation}
\begin{aligned}
 e_r^i = - i \pi (k_0/ \eta^2) u_\varphi^i,  \quad h_r^i = i \pi (k_0/ \eta^2) \sqrt {\epsilon_0 / \mu_0} s_\varphi^i. 
\end{aligned} \nonumber
\end{equation}
Equations (\ref{eq:S_theta}) - (\ref{eq:U_phi}) are written in a general form. {  The formalism does not pose any restrictions on the position or nature of the excitation sources, meaning structured light optical vortices, vector beams \cite{moreno2010decomposition} and optical skyrmions can serve as inputs \cite{shen2025free,chen2025more,wang2024topological}. What is needed  for a given input field is the expansion of the incident field and the scattered fields in the surrounding space (i.e., Region I) as well as inside the magnetically ordered materials  in terms of the same basis which allows the boundary conditions on the surfaces of the sphere to be matched. More details are provided in Appendix C.} Without loss of generality in this work we analyze only the lowest-order spherical waves generated by elementary sources as shown in (C1) and (C2) of Appendix C. To emphasize the generality  of the formalism, in Appendix C we briefly describe also an implementation of radially and azimuthally polarized beams. It is established  that the sum of the radially and azimuthally polarized beams produce a circularly polarized vortex beam which  is  used to probe the magnetic vortex dynamics \cite{luttmann2025optical}. 

In our previous works for periodic planar geometries \cite{jandieri1,jandieri2,jandieri3} we expanded the fields into a set of space
harmonics varying as $e^{jk_{xn}x}$, where $k_{xn}$ are the wavenumbers along the periodicity (along the $x$-axis). Here, however, we are dealing with the spherical symmetry and therefore, we need to use the fields' expansion in terms of the polar $\theta$ angle (meaning  in the basis of $e^{ji\pi\kappa}$).      

At the singularity, the magnetic moment and hence the scattering from it vanishes.  To deal with this region numerically, we define the fields at $r=0$ that appear in the denominator of the block matrices in Appendix B, and a small homogeneous spherical region surrounding the global origin and having a radius $b$ is considered. It is denoted as Region II (c.f. inset in Fig. \ref{fig::1}) and  without loss of generality is assumed to be the dielectric background space. The fields in Region II can be written in a similar form as those in Region I. The main difference is that instead of the Schelkunoff Hankel function $\hat{h}^{(2)}_n(\eta)$ we will have a standing wave that is expressed by the Schelkunoff Bessel function $\hat{j}_n(\eta)$. This is because an incoming wave $\hat{h}^{(1)}_n(\eta)$ is completely reflected at the global origin thus generating $\hat{h}^{(2)}_n(\eta)$, which leads to the formation of the total field  - the standing wave $\hat{j}_n(\eta) = 0.5[\hat{h}^{(1)}_n(\eta)+\hat{h}^{(2)}_n(\eta)]$ - in Region II \cite{jandyasu}. This significantly differs from the analysis for the planar geometry \cite{jandieri1,jandieri2, khomeriki}, where the transmitted field in the innermost region is represented in terms of incoming plane waves. Moreover, there are no incident fields in Region II and hence, the last terms on the right side in (\ref{eq:S_theta})-(\ref{eq:U_phi}) should be set to zero. 

Finally, by matching the boundary conditions at $r=a$ and $r=b$, we obtain  the following two equations from which the unknown $\boldsymbol{C}^{(1)}$ can be  determined 
\begin{align}
&\left[ \overline{\overline{\boldsymbol{P}}}_U^{(1)} - \begin{pmatrix} \mathbf{K} & 0 \\ 0 & \tilde{\mathbf{K}} \end{pmatrix} \begin{pmatrix} 0 & \mathbf{K} \\ -\tilde{\mathbf{K}} & 0 \end{pmatrix}^{-1} \overline{\overline{\boldsymbol{P}}}_S^{(1)} \right] \boldsymbol{C}^{(1)} \nonumber =\\
&= \begin{pmatrix} \boldsymbol{U}_{\theta}^{\text{inc}} \\ \boldsymbol{U}_{\varphi}^{\text{inc}} \end{pmatrix} - \begin{pmatrix} \mathbf{K} & 0 \\ 0 & \tilde{\mathbf{K}} \end{pmatrix} \begin{pmatrix} 0 & \mathbf{K} \\ -\tilde{\mathbf{K}} & 0 \end{pmatrix}^{-1} \begin{pmatrix} \boldsymbol{S}_{\theta}^{\textit{inc}} \\ \boldsymbol{S}_{\varphi}^{\textit{inc}} \end{pmatrix} \label{eq:19}
\\
&\left[ \overline{\overline{\boldsymbol{P}}}_U^{(L)} - \begin{pmatrix} \mathbf{G} & 0 \\ 0 & \tilde{\mathbf{G}} \end{pmatrix} \begin{pmatrix} 0 & \mathbf{G} \\ -\tilde{\mathbf{G}} & 0 \end{pmatrix}^{-1} \overline{\overline{\boldsymbol{P}}}_S^{(L)} \right] \boldsymbol{C}^{(L)} = 0 \label{eq:20}
\end{align}
with
\begin{equation}
\begin{aligned}
&\mathbf{G} = \{G_{in}\} = j \mathrm{g}_{in} \hat{j}^{'}_n (\eta), 
&\tilde{\mathbf{G}} = \{\tilde{{G}}_{in}\} = \hat{\mathrm{g}}_{in} \hat{j}_n (\eta) .
\end{aligned}
\end{equation}
Once $\boldsymbol{C}^{(1)}$ is found, the fields  inside (using (\ref{eq::9}) and (\ref{eq::10}) together with (\ref{eq::13})), as well as outside (using (\ref{eq:S_theta}) - (\ref{eq:18})) the BP are fully determined.

\begin{figure}[t]
\includegraphics[width=\linewidth]{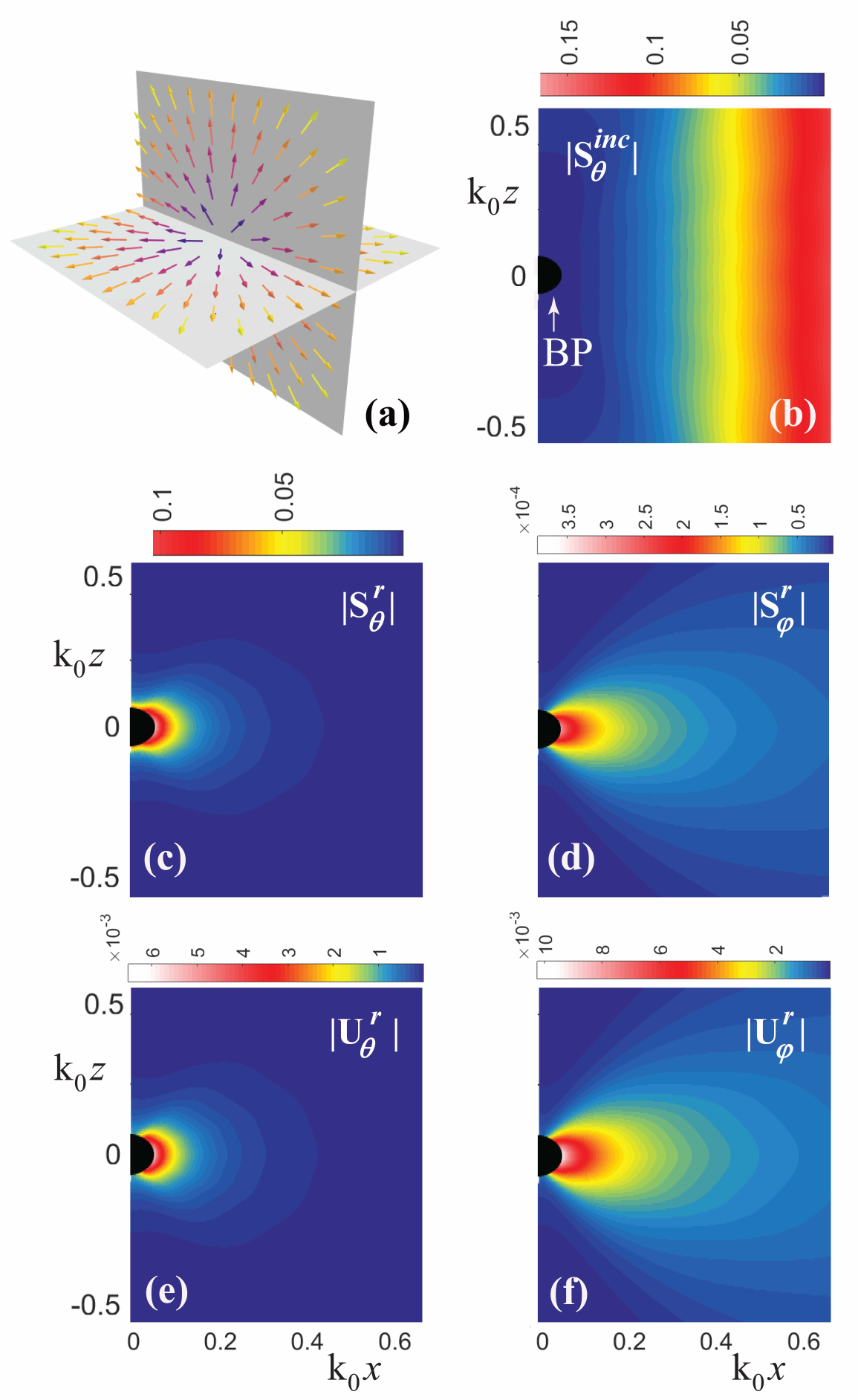}
\caption{(a) The hedgehog Bloch point ($\gamma = 0^0$); (b) the incident  electric fields $|S^{inc}_\theta|$; (c), (d) the reflected electric fields $|S_\theta^{r}|$ and $|S_\varphi^{r}|$, respectively; (e),(f) the reflected magnetic fields $|U_\theta^{r}|$ and $|U_\varphi^{r}|$, respectively. Here $k_0 a = 0.06$, $k_0 b = 0.005$, $\epsilon_r = 12$ and $k_0 h = 4$.}\label{fig::2}
\end{figure}

\begin{figure}[t]
\includegraphics[width=\linewidth]{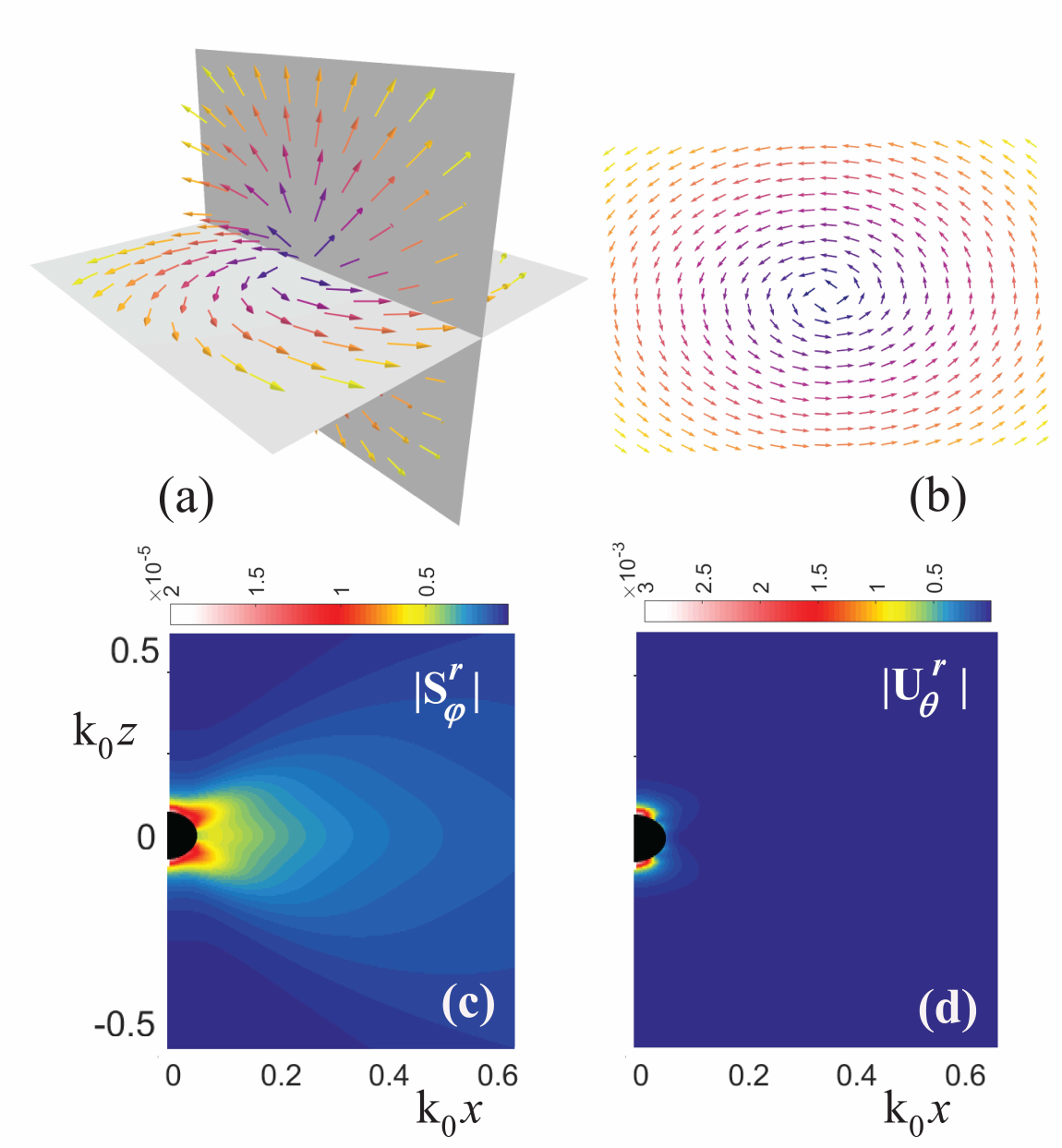}
\caption{(a) The twisted BP; (b) its projection on the equatorial cut ($\theta = 90^0$); (c),(d) the spatial distributions of the reflected electric $|S_\varphi^{r}|$ and magnetic $|U_\theta^{r}|$ fields, respectively. The other parameters are the same as those in Fig. \ref{fig::2}.}
\label{fig::3}
\end{figure}

\begin{figure}[t]
\includegraphics[width=\linewidth]{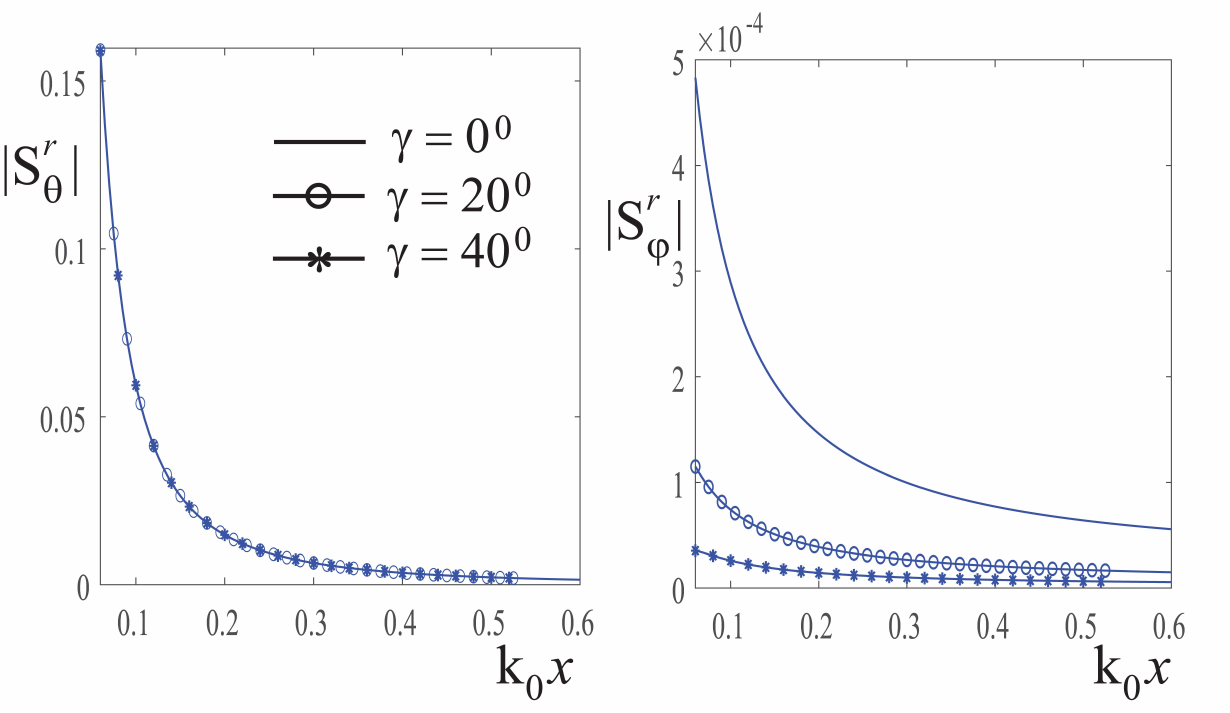}
\caption{ The reflected electric field components $|S_\theta^{r}|$ (left figure) and $|S_\varphi^{r}|$ (right figure) versus $k_0x$ at a fixed polar angle $\theta = 90^0$ for different azimuthal rotation angles $\gamma$: $\gamma = 0^0$ (solid line); $\gamma = 20^0$ (circles);  $\gamma = 40^0$ (asterisk). The other parameters are the same as those in Fig. \ref{fig::2}. }\label{fig::4}
\end{figure}

\section{Numerical Results and Discussions}
%CAN YOU PLOT FOR FIG.1 THE EXTERNAL FIELD VECTOR AS THEY ACT LOCALLY ON THE MAGNETIC MOMENTS; SAY IN THE EQUATORIAL PLANE

To validate the correctness and  accuracy of the proposed formalism we begin by studying  two relatively simple models of electromagnetic scattering of the spherical transverse magnetic $TM_r$ and transverse electric $TE_r$ incident waves on: a) a  homogeneous dielectric sphere, and b) a dielectric sphere with radially varying relative dielectric permittivities. The results are then compared with analytical solutions obtained based on the 3-D transition matrix (T-matrix) theory \cite{jandyasu} which utilizes the expressions for the fields in terms of spherical functions in all regions. The results are not shown here for brevity, however, very good agreement is found   for all cases. Note that for a homogeneous scatterer the fields are decoupled and only a set of two differential equations for ($E_r, S_\theta, U_\varphi$) corresponding to the $TM_r$ wave and ($H_r, U_\theta, S_\varphi$) corresponding to the $TE_r$ wave are solved separately. This is not the case for the analysis of the BPs, which is characterized by the magnetic permeability matrix. Therefore, a set of four coupled differential equations have to be solved, as shown in Appendix B. 
\subsubsection{Hedgehog magnetic texture}
As the simplest interesting example, {  consider an electric dipole} radiation source  positioned  on  the  $z$-axis and pointing along it (see Fig.\ref{fig::1}). This source radiates the lowest-order spherical waves expressed by a zero-th order spherical Hankel function \cite{2823} (see Appendix C). {  The only non-zero components of the incident field produced by this elementary dipole are $U^{inc}_\varphi$, $S^{inc}_r$ and  $S^{inc}_\theta$ (these field components are derived from the expression for the radially directed magnetic vector potential). Hence, the second terms on the right-hand side in (\ref{eq:S_phi}) and (\ref{eq:U_theta}) are set to zero. Although our formalism is general, in this paper we consider a relatively simple model, when there is not a magnetic material beyond the BP and the source is located in the free space surrounding the BP. The goal of this paper is to capture the physics of BP and present fundamental studies about its interaction with the incoming spherical wave, i.e. to study how the fields are modified by the singular magnetic texture. The present approach allows, in principle, to study bulk and interfaces with regard to dielectric as well as to background-magnetic properties.} 

Firstly, a hedgehog Bloch point ($\gamma=0^0$), when the unit vectors of magnetization inside the BP have only radial components ($m_r^0 \neq 0$, $m_\theta = m_\varphi =0$), is considered (c.f. Fig. \ref{fig::2}(a)).  This structure can be viewed as an emergent magnetic monopole. After several tests for convergence of the solutions, the truncation number was chosen as $N = 5$. The BP is divided into 25 equal thin shell layers and all material parameters are assumed to be constant within each thin shell layer (see Sec. II). The absolute value of the incident electric field $|S^{inc}_\theta|$ and the reflected electric  $|S_\theta^{r}|$,  $|S_\varphi^{r}|$ 
%and the time-averaged optical chiral density $<\chi>$ 
and magnetic $|U_\theta^{r}|$,  $|U_\varphi^{r}|$ fields are demonstrated in Figs. \ref{fig::2}(b) - \ref{fig::2}(f) at $k_0 a = 0.06$, $k_0 b = 0.005$, $\epsilon_r = 12$ and $k_0 h = 4$. {  In this work we develop a general, material-independent formulation, and do not specify a particular material. However, we should note that the radius of the sphere $a$ should be taken to be much smaller than the wavelength, since the electromagnetic waves are in the range of magnetic excitations.} The formalism analytically and numerically is not restricted to a particular location of the source. {  Here, the dipole  is chosen to be located at a relatively large distance from the BP and hence, the incident wavefront impinging on the BP has a symmetric profile with respect to the polar $\theta$ angle (for a dipole source closer to the BP, the field would lose its symmetric profile and would be concentrated in the vicinity of the upper half of the sphere).} The reflected fields show typical  profiles with  strong localization  in the vicinity of the BP at around $\theta = 90^0$ and the $\theta$-components of both fields decrease  faster than the $\varphi$-components with increasing a distance from the surface of the BPs.

\begin{figure}[t]
\includegraphics[width=\linewidth]{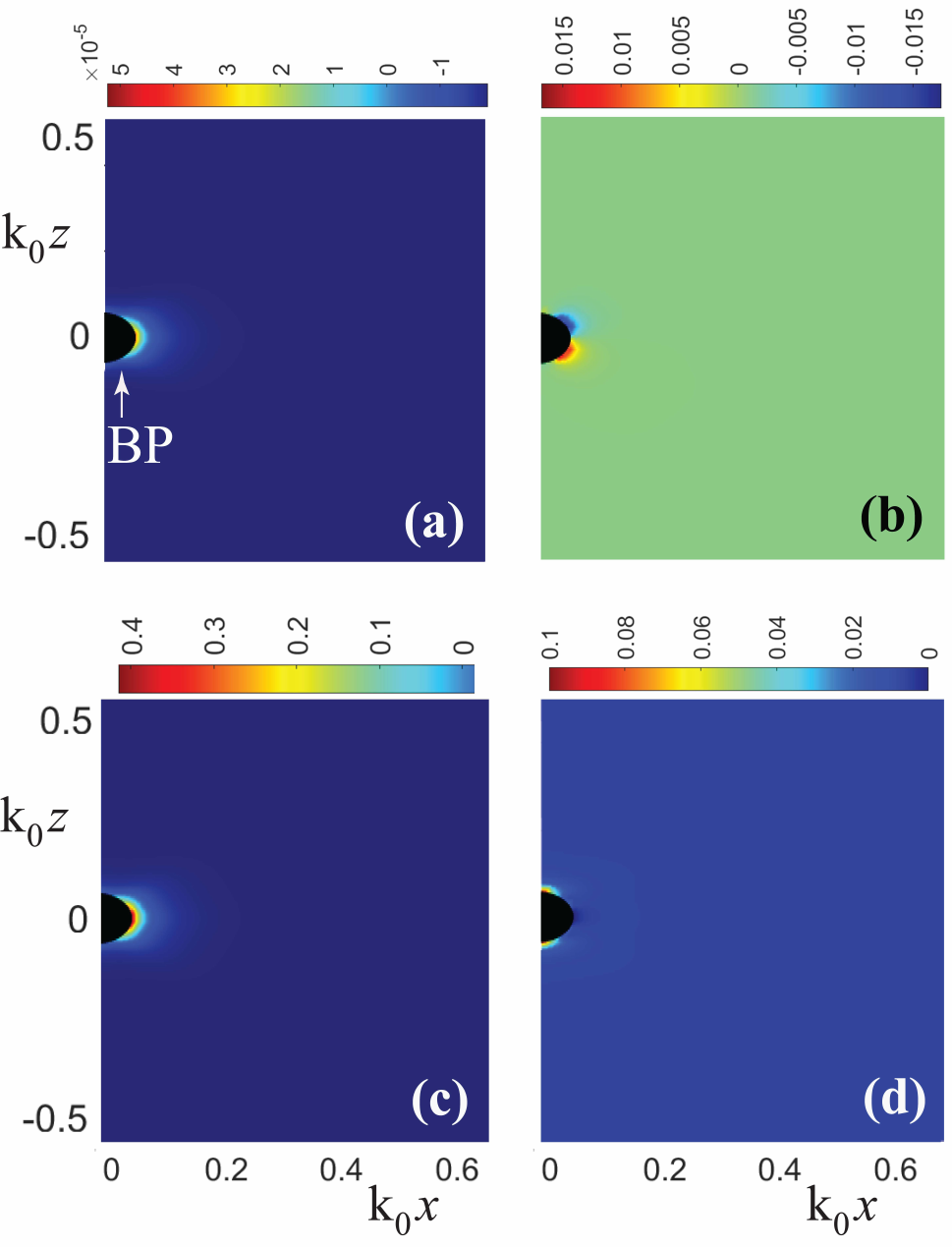}
\caption{ (a), (b): Time-averaged gauge-invariant optical chiral density and (c),(d):  the magnetoelectric density for the hedgehog Bloch point (left figures) and the twisted Bloch point $\gamma = 60^0$ (right figures). The other parameters are the same as in Fig. \ref{fig::2}. }
\label{fig::5}
\end{figure}

\begin{figure}[t]
\includegraphics[width=\linewidth]{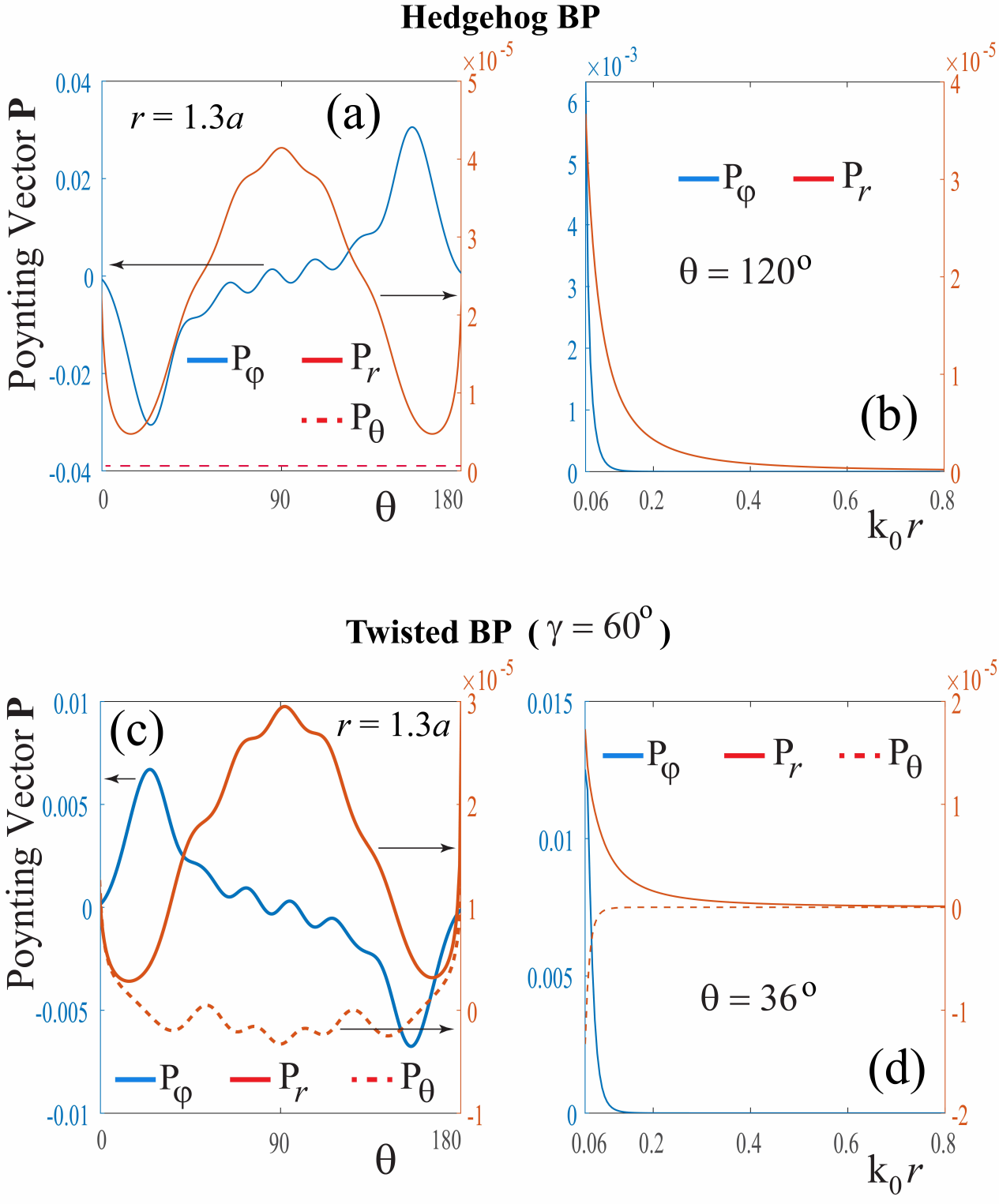}
\caption{ Time-averaged Poynting vector of the reflected fields as a function of (a): the polar angle $\theta$ at the fixed radial distance $r=1.3a$, i.e. very close to the outer surface of the hedgehog BP; (b) the normalized radial distance $k_0 r$ at the spherical cut $\theta=120^0$. (c),(d): the same as above but for the twisted BP at $\gamma = 60^0$. }   
\label{fig::6}
\end{figure}

\begin{figure}[t]
\includegraphics[width=\linewidth]{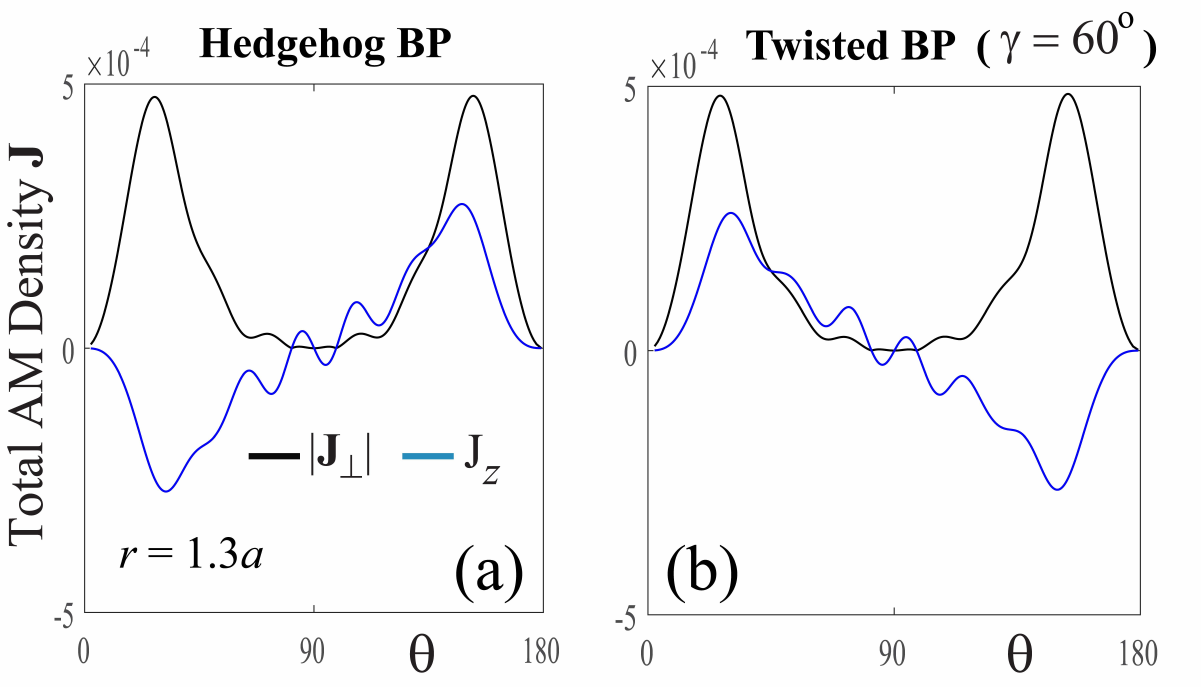}
\caption{ Total angular momentum (AM) density $|J_ \perp|$ (black line) and $J_z$ (blue line) as a function of the polar angle $\theta$ at the fixed radial distance $r=1.3a$ for: (a) the hedgehog BP; (b) the twisted BP. } \label{fig::7}
\end{figure} 

\subsubsection{Twisted Bloch point}
Let us consider the effect of the azimuthal rotation angle $\gamma$ generating twisted BP in Fig. \ref{fig::3}(a) on the EM wave scattering. Unlike the hedgehog BP, in the case of the twisted BP at $\gamma \neq 0$, the unit vector of magnetization is radially oriented only at the poles of $\theta = 0^0$ and $\theta = 180^0$. Its orientation as a function of $\theta$ is defined by (\ref{eq:5}). Let us inspect   the field distributions of $|U_\theta^r|$ and $|S_\varphi^r|$ at $\gamma = 60^0$. These field components  are  excited solely by an interaction of the incident field with the BP and contribute to a formation of the scattered fields, in particular they are not present in the incident field, i.e., $S_\varphi ^{inc} = U_\theta ^{inc} = 0$. The results of the field distributions are given in Figs. \ref{fig::3}(b) and 3(c). From the figures it follows that the field distributions for $|S_\varphi^r|$ and $|U_\theta^r|$ exhibit quite different profiles from those in Figs. \ref{fig::2}(d) and \ref{fig::2}(e). The fields are no longer localized at around $\theta = 90^0$, but are concentrated closer to the $z$-axis of the twisted BP. The results are not shown, but
interestingly, $S_\theta^r$ and $U_\varphi^r$ do not show any changes from those in Figs. \ref{fig::3}(c) and 3(f) at $\gamma = 0^0$. To study this phenomenon in more detail,  $|S_\theta^r|$ and  $|S_\varphi^r|$ are plotted as a function of $k_0x$ at a fixed spherical angle $\theta = 90^0$ for different values of $\gamma$:  $\gamma = 0^0$ (solid line); $\gamma = 20^0$ (circles);  $\gamma = 40^0$ (asterisk). The results of the reflected electric fields are illustrated in Fig. 4. As expected,  $S_\theta^r$ (black line) does not show any changes when increasing $\gamma$, whereas a maximum of $S_\varphi^r$ (red line) is decreasing inversely proportional to $\gamma$. The results for the magnetic field show  similar profiles.
Namely, $U_\varphi^r$ is not changing by increasing $\gamma$. 
\subsubsection{Emergent features of formed EM fields and magnetoelectric coupling}
Having shown how the noncollinear magnetic texture mold the incoming EM waves, it is of interest to consider the specific features of these fields. Here, several well-established quantities will be discussed. 
{  The time-average of the important magnetoelectric  pseudoscalar  $\mathbf{E}\cdot\mathbf{H}$ may be expressed as
$
Q_{ME}= \mathbf{E}\cdot\mathbf{H} = \frac{1}{2}\left[S_\theta U_\theta^*+\frac{S_\varphi U_\varphi^*}{1-\kappa^2}-
\frac{1}{\eta^2}\frac{\partial U_\varphi}{\partial \kappa} \left(\frac{\partial S_\varphi}{\partial \kappa}\right)^* 
\right]$
from which the time-averaged,  normalized (propagating) optical chiral density $\chi$ follows as 
\cite{jandieri2024chiral}: 
\begin{equation}
\begin{aligned}
\chi = \frac{1}{\eta^2}
\Im\left(Q_{ME}\right), 
\end{aligned}  \label{eq:23}
\end{equation}
and the reactive  magnetoelectric density $A_{ME}$ is \cite{bliokh2014magnetoelectric} $A_{ME} = \frac{1}{\eta^2}
\Re\left(Q_{ME}\right).$ 
$Q_{ME}$ is derived directly from the fields and hence $\chi$ and $A_{ME}$ are gauge-invariant. }These values are normalized by $\epsilon_0 k_0^3/2$. Here, we recall the importance of
producing fields with time-averaged, finite and  possibly tunable $\mathbf{E}\cdot\mathbf{H}$ for research concerning chiral materials \cite{PhysRevLett.104.163901}, axion fields in general,  and axion electrodynamics
of a 3-D topological insulator in particular as well as to access spin-charge  response in condensed matter \cite{Li2010,doi:10.1126/science.aaf5541,Nenno2020,PhysRevB.99.075104}.
Our results for $\chi$ and $A_{ME}$ in Fig. \ref{fig::5} for  hedgehog and twisted BPs are generated by  fields
determined using (\ref{eq:7}). Irradiating with  fields having no $\chi$ and $A_{ME}$, meaning any transverse fields with real-valued $\mathbf{E}$,  upon interaction with the magnetic texture,
the phases and the directions of $\mathbf{E}$ and $\mathbf{H}$ are mixed leading to a formation of localized finite $\chi$ and $A_{ME}$. The twist in a BP allows for more channels for scattering and hence more mixing enhancing the maximal magnitude of $\chi$ and $A_{ME}$.
The strong localization  of the produced magnetoelectric activity implies  that if we have a collection of BP-type textures, as they occur for example in artificial spin ice or gyroid structures \cite{doi:10.1021/acsami.4c02366}, then  a lattice of $\chi$ and $A_{ME}$ will be formed. 
Notably,  $\mathbf{E}\cdot\mathbf{H}$ has been observed experimentally in the center  of  plasmonic vortices \cite{PhysRevResearch.6.013163}. The method  identified here generates finite, time averaged $\chi$ and $A_{ME}$ in a non-dissipative way and they are localized well beyond the BPs, whereas  in plasmonic-based methods $\chi$ and $A_{ME}$  live at the metal-dielectric interface. Thus, our scheme is qualitatively different and complementary to the plasmonic method. This remark becomes  relevant when using $\chi$ and $A_{ME}$ to probe  unperturbed samples.

For optical forces and torques as well as  for an insight into the energy flow and the orbital angular momentum distribution, the components of the time-averaged Poynting vector $\mathbf{P}$ are important. They are calculated according to the following expressions:
\begin{equation}
\begin{aligned}
P_r &= \frac{1}{2\eta^2\sqrt{1-\kappa^2}}\Re(S_\theta U_\varphi^*-S_\varphi U_\theta^*) ,\\
P_\theta &= \frac{1}{2\eta^3\sqrt{1-\kappa^2}}\Re\left(\frac{U_\varphi^*}{j}\frac{\partial U_\varphi}{\partial \kappa} + S_\varphi\left(\frac{1}{j}\frac{\partial S_\varphi}{\partial \kappa}\right)^* \right),\\
P_\varphi &= \frac{1}{2\eta^3\sqrt{1-\kappa^2}}\Re\left(-\frac{U_\theta^*}{j}\frac{\partial U_\varphi}{\partial \kappa} - S_\theta\left(\frac{1}{j}\frac{\partial S_\varphi}{\partial \kappa}\right)^* \right).
\end{aligned}
\label{eq:24}
\end{equation}
These components are shown  in Figs. \ref{fig::6}(a) and \ref{fig::6}(c) for the hedgehog BP and the twisted BP  as a function of $\theta$ at a fixed radial distance $r=1.3a$  close to the outer surface. The Poynting vector components are normalized by $k_0^2 \sqrt {\epsilon_0 / \mu_0}$. The main contribution is carried by  $P_\varphi$ (blue line) which is antisymmetric with respect to the equatorial cut $\theta = 90^0$. We recall that  $P_\varphi$ is the only  component of the Poynting vector which is predominantly influenced by an interaction of the electromagnetic wave with the BP (as inferred from the third expression in (\ref{eq:24}) which indicates that   $P_\varphi = 0$ if the magneto-optical activity of the BP disappears).  Further  numerical results and analysis  evidence that the shape of $P_\varphi$, as expected, does not show any substantial variations with respect to the radial distance  $r$ (in particular its antisymmetric profile is not changing). Another point  worth noticing is that unlike the hedgehog BP, the $\theta$ component of the Poynting vector is  more pronounced for the twisted BP (red dashed line). As for the radial component of the Poynting vector $P_r$,  similar profiles (red line) are exhibited for both types of BPs.
%and it is symmetric with respect to $\theta = 90^0$. 
Note  that the radial component of the Poynting vector is mainly affected by the homogeneous dielectric permittivity of the scatterer and is not    sensitive to the elements of the magnetic permeability matrix that characterize the BP. 
Assuming $g \mu_0 M_s =0 $ one readily concludes that  all non-diagonal components of the permeability matrix in Appendix A become zero and the diagonal components are equal to unity. Dependence of the Poynting vector components versus the normalized radial distance $k_0 r$ in some particular spherical cuts is shown in Figs. \ref{fig::6}(b) and \ref{fig::6}(d). 
The $P_\varphi$ and $P_\theta$ components are rapidly decreasing in magnitudes as the radial distance grows but are localized clearly outside the BPs. In the far-field region (with respect to the BP size) only the radial component of the Poynting vector survives, as expected for EM wave scattering by homogeneous objects.   

Generally, the  total angular momentum density follows from $
    \mathbf{J} = \mathbf{r} \times \mathbf{P}/c^2 $. For strongly non-paraxial beams, it is not unique how the 
orbital angular momentum density $\mathbf{L}$ can be disentangled from the spin angular momentum density $\mathbf{S}$ as is done in \cite{PhysRevA.82.063825} for paraxial beams ($\mathbf{L}= \mathbf{J} - \mathbf{S}$). Hence, we will discuss here $\mathbf{J}$, keeping in mind that it encompasses both $\mathbf{L}$ and $\mathbf{S}$. The transversal $|\boldsymbol{J}_ \perp|=\sqrt{J_x^2+J_y^2}$ and the longitudinal $J_z$ components of the total angular momentum (AM) density are given by 
\begin{equation}
\begin{aligned}
|\boldsymbol{J}_ \perp|=\sqrt{\kappa^2J_\theta^2+J_\varphi^2}, \quad J_z = -\sqrt{1-\kappa^2}J_\theta
\end{aligned}  \label{eq:25}
\end{equation}
with $J_\theta = -rP_\varphi$ and $J_\varphi=rP_\theta$, where $r$ is taken with respect to the global origin. 
The total AM density $|\boldsymbol{J}_ \perp|$ (black line) and $J_z$ (blue line) as a function of the polar angle $\theta$ at the fixed radial distance $r=1.3a$ for the hedgehog BP and for the twisted BP are shown in Figs. \ref{fig::7}(a) and 7(b), respectively. Obviously, in the absence of the magneto-optical effects characterizing the BP, $P_\varphi = 0$, i.e. $J_\theta = 0$ and hence, no total AM density along the $z$-axis is observed. As for the total AM, i.e. an integration with respect to $r$ and $\theta$, under our prescribed geometrical parameters and the incident field, it will be equal to zero. For a source  closer to the BP, the reflected field is substantially  stronger, concentrated in the vicinity of the upper half of a sphere enclosing the BP, and a finite total AM results. {  Finally, we note that some experimental studies for the chiral fields and orbital angular momentum-dependent responses are described in \cite{mun2020electromagnetic}. They can be useful for the analysis of the reflected fields by the BP discussed in this paper, which possess orbital angular momentum, chirality as well as magnetoelectric densities.}
%PhysRevLett.134.156701
\section{Conclusions}
The work presents an efficient and self-consistent approach for the analysis of EM wave scattering by  hedgehog and twisted BPs. The goal was to investigate how   non-collinear magnetic textures respond to EM fields in the frequency range of magnetic excitations and how to use   BPs as photonic elements that modify the scattered (reflected) fields. The formalism is  general and can be applied to  various excitation sources and spin configurations apart from  BPs. Special attention is paid to   field components generated solely by an interaction of   EM waves and   BPs and to their modifications when transforming from  a hedgehog   to a twisted configuration.  The field's spatial distributions,  chirality density, magnetoelectric density, Poynting vector, and total angular momentum density are numerically evaluated and analyzed. 

The generalization of the formalism to the problem with both $\theta$ and $\varphi$-dependent spherical objects is straightforward and requires developing  explicit expressions for the field components (including the ${\partial }/{\partial \varphi}$ terms) in a spherical coordinate system satisfying the vector Helmholtz equation. As a result,
%a modification of Appendix B and also Eqs.(\ref{eq:S_theta}) - (\ref{eq:U_phi}) is required.  The 
the size of the matrix to be solved becomes  larger than in the problem considered here, but the analytical procedure remains the same.
Finally, it is important to mention that our EM simulations are carried out for a particular size of the BP, the extent to which depends on the temperature  $T$ \cite{YASTREMSKY2025172834}.
For a fixed temperature, for example, near the Curie temperature $T_c$ the BP radius scales as $\propto (T-T_c)^{-1/2}$.

\section*{Author contributions}
All authors equally contributed to this work. 

\section*{Acknowledgments}
The work has been supported by the Shota Rustaveli National Science Foundation of Georgia (SRNSFG) [FR-23-249] and  the DFG under project Nr. 429194455,  and DOE (USA) Nr.~DE-FOA-0002514. D.E. and V.J. also acknowledge the partial support by the Deutsche Forschungsgemeinschaft (DFG) in the Framework of the CRC/TRR 196 MARIE (Project-ID 287022738) within project M03.

V. Jandieri would like to express his sincere gratitude to
Professor Kiyotoshi Yasumoto from Kyushu University, Japan
for discussions about the theory of periodic and bandgap structures.

%\section*{Financial disclosure}

%None reported.

\section*{Conflict of interest}

The authors declare no potential conflicts of interest.

\section*{Data Availability Statement}

The data beyond those in the text   are available from the  authors upon reasonable request.

%\subsection*{Appendix A}
\appendix
\section{Magnetic Permeability Matrix}
The elements of the magnetic permeability matrix:
%\begin{equation}
$
\mu_{ij}(\theta, r) =
\begin{bmatrix}
\mu_{11} & \mu_{12} & \mu_{13} \\
\mu_{12}^* & \mu_{22} & \mu_{23} \\
\mu_{13}^* & \mu_{23}^* & \mu_{33}
\end{bmatrix}
$
%\end{equation}
are implicitly written in the following form:
%\begin{widetext}
%\begin{equation}
%\begin{aligned}
%\mu_{11} &= 1 + \alpha \dfrac{\mathrm{g}^2 \mu_0^2 M_s^2}{\omega^2}
%\left((m_\theta^0)^2 + (m_\phi^0)^2\right)\left[\hat{O}M_0(r)\right]
%\end{aligned}
%\end{equation}
%\begin{equation}
%\begin{aligned}
%\mu_{12} &= \alpha\left(
%    -\dfrac{\mathrm{g}^2 \mu_0^2 M_s^2}{\omega^2} m_r^0 m_\theta^0 \left[\hat{O}M_0(r)\right]
%    - j \dfrac{g \mu_0 M_s}{\omega} m_\phi^0 \right)
%\end{aligned}
%\end{equation}
%\begin{equation}
%\begin{aligned}
%\mu_{13} &= \alpha\left(
%    -\dfrac{\mathrm{g}^2 \mu_0^2 M_s^2}{\omega^2} m_r^0 m_\phi^0 \left[\hat{O}M_0(r)\right]
%    + j \dfrac{g \mu_0 M_s}{\omega} m_\theta^0 \right)
%\end{aligned}
%\end{equation}
%\begin{equation}
%\begin{aligned}
%\mu_{22} &= 1 + \alpha \dfrac{\mathrm{g}^2 \mu_0^2 M_s^2}{\omega^2}
%\left((m_r^0)^2 + (m_\phi^0)^2\right)\left[\hat{O}M_0(r)\right]
%\end{aligned}
%\end{equation}
%\begin{equation}
%\begin{aligned}
%\mu_{23} &= \alpha\left(
%    -\dfrac{\mathrm{g}^2 \mu_0^2 M_s^2}{\omega^2} m_\theta^0 m_\phi^0 \left[\hat{O}M_0(r)\right]
%    - j \dfrac{g \mu_0 M_s}{\omega} m_r^0 \right)
%\end{aligned}
%\end{equation}
%\begin{equation}
%\begin{aligned}
%\mu_{33} &= 1 + \alpha \dfrac{\mathrm{g}^2 \mu_0^2 M_s^2}{\omega^2}
%\left((m_r^0)^2 + (m_\theta^0)^2\right)\left[\hat{O}M_0(r)\right]
%\end{aligned}
%\end{equation}
%\begin{equation}
%\begin{aligned}
%\mu_{21} &= \mu_{12}^{*}, \qquad
%\mu_{31} = \mu_{13}^{*}, \qquad
%\mu_{32} = \mu_{23}^{*}
%\end{aligned}
%\end{equation}
\begin{widetext}
\begin{equation}
\mu_{11} = 1 + \alpha \dfrac{\mathrm{g}^2 \mu_0^2 M_s^2}{\omega^2}
\left((m_\theta^0)^2 + (m_\phi^0)^2\right)\left[\hat{O}M_0(r)\right],\; 
\mu_{12} = \alpha\left(
    -\dfrac{\mathrm{g}^2 \mu_0^2 M_s^2}{\omega^2} m_r^0 m_\theta^0 \left[\hat{O}M_0(r)\right]
    - j \dfrac{g \mu_0 M_s}{\omega} m_\phi^0 \right),
    \end{equation}
    \begin{equation}
\mu_{13} = \alpha\left(
    -\dfrac{\mathrm{g}^2 \mu_0^2 M_s^2}{\omega^2} m_r^0 m_\phi^0 \left[\hat{O}M_0(r)\right]
    + j \dfrac{g \mu_0 M_s}{\omega} m_\theta^0 \right),\;  
\mu_{22} = 1 + \alpha \dfrac{\mathrm{g}^2 \mu_0^2 M_s^2}{\omega^2}
\left((m_r^0)^2 + (m_\phi^0)^2\right)\left[\hat{O}M_0(r)\right],\; 
 \end{equation}
    \begin{equation}  
\mu_{23} = \alpha\left(
    -\dfrac{\mathrm{g}^2 \mu_0^2 M_s^2}{\omega^2} m_\theta^0 m_\phi^0 \left[\hat{O}M_0(r)\right]
    - j \dfrac{g \mu_0 M_s}{\omega} m_r^0 \right),\; 
\mu_{33} = 1 + \alpha \dfrac{\mathrm{g}^2 \mu_0^2 M_s^2}{\omega^2}
\left((m_r^0)^2 + (m_\theta^0)^2\right)\left[\hat{O}M_0(r)\right],\; 
\end{equation}
    \begin{equation}  
\mu_{21} = \mu_{12}^{*}, \qquad
\mu_{31} = \mu_{13}^{*}, \qquad
\mu_{32} = \mu_{23}^{*}
\end{equation}

% \begin{align}
% \mu_{11} &= 1 + \alpha \dfrac{\mathrm{g}^2 \mu_0^2 M_s^2}{\omega^2}
% \left((m_\theta^0)^2 + (m_\phi^0)^2\right)\left[\hat{O}M_0(r)\right] &\\
% \mu_{12} &= \alpha\left(
%     -\dfrac{\mathrm{g}^2 \mu_0^2 M_s^2}{\omega^2} m_r^0 m_\theta^0 \left[\hat{O}M_0(r)\right]
%     - j \dfrac{g \mu_0 M_s}{\omega} m_\phi^0 \right) &\\
% \mu_{13} &= \alpha\left(
%     -\dfrac{\mathrm{g}^2 \mu_0^2 M_s^2}{\omega^2} m_r^0 m_\phi^0 \left[\hat{O}M_0(r)\right]
%     + j \dfrac{g \mu_0 M_s}{\omega} m_\theta^0 \right) \\
% \mu_{22} &= 1 + \alpha \dfrac{\mathrm{g}^2 \mu_0^2 M_s^2}{\omega^2}
% \left((m_r^0)^2 + (m_\phi^0)^2\right)\left[\hat{O}M_0(r)\right] \\
% \mu_{23} &= \alpha\left(
%     -\dfrac{\mathrm{g}^2 \mu_0^2 M_s^2}{\omega^2} m_\theta^0 m_\phi^0 \left[\hat{O}M_0(r)\right]
%     - j \dfrac{g \mu_0 M_s}{\omega} m_r^0 \right) \\
% \mu_{33} &= 1 + \alpha \dfrac{\mathrm{g}^2 \mu_0^2 M_s^2}{\omega^2}
% \left((m_r^0)^2 + (m_\theta^0)^2\right)\left[\hat{O}M_0(r)\right] \\
% \mu_{21} &= \mu_{12}^{*}, \qquad
% \mu_{31} = \mu_{13}^{*}, \qquad
% \mu_{32} = \mu_{23}^{*}
% \end{align}
where
\begin{equation}
    \alpha = \dfrac{1}{1 - \dfrac{\mathrm{g}^2 \mu_0^2 M_s^2}{\omega^2} \left[\hat{O}M_0(r)\right]^2},\; 
    \hat{O} = \left(\frac{\partial^2}{\partial r^2} + \frac{2}{r} \frac{\partial}{\partial r} - \frac{2}{r^2}\right) + 2\nu \left(1 - M_0^2\right)
\end{equation}
Here $\delta_{ij}$ is the Kronecker delta function, the asterisk denotes a complex conjugate and $\omega$ is an excitation angular frequency.

\section{Block Matrices in Spherical Coupled-Wave Analysis}

Expressions of the block matrices  ${\Psi_{qq'}}$ in (\ref{eq:8}) are explicitly written as:
%\begin{widetext}
  %
\begin{equation}
    \Psi_{11} = 0,\; 
    \Psi_{12} = -\frac{1}{j\eta} \llbracket \mu_{13}^* \rrbracket \llbracket \mu_{11} \rrbracket^{-1} \llbracket j\pi p \rrbracket ,
    \Psi_{13} = -\left( \llbracket \mu_{23}^* \rrbracket 
    - \llbracket \mu_{13}^* \rrbracket \llbracket \mu_{11} \rrbracket^{-1} \llbracket \mu_{12} \rrbracket \right) 
    ,
    \end{equation}
    \begin{equation}
     \Psi_{14} = -\frac{1}{\eta^2} \llbracket \sqrt{1 - \kappa^2} \rrbracket \llbracket j\pi p \rrbracket \left\llbracket 
     \frac{1}{\epsilon_r} \right\rrbracket \llbracket j\pi p \rrbracket  - 
    \left\llbracket \frac{\mu_{33}}{\sqrt{1 - \kappa^2}} \right\rrbracket
    + \llbracket \mu_{13}^* \rrbracket \llbracket \mu_{11} \rrbracket^{-1} \left\llbracket \frac{\mu_{13}}{\sqrt{1 - \kappa^2}} \right\rrbracket ,\; 
    \Psi_{21} = 0,
    \end{equation}
    \begin{equation}
    \Psi_{22} = \frac{1}{j\eta} \left\llbracket \frac{1}{\sqrt{1 - \kappa^2}} \right\rrbracket^{-1} 
    \llbracket \mu_{12}^* \rrbracket \llbracket \mu_{11} \rrbracket^{-1} \llbracket j\pi p \rrbracket ,\;
    \Psi_{23} = \left\llbracket \frac{1}{\sqrt{1 - \kappa^2}} \right\rrbracket^{-1} 
    \left( \left\llbracket \mu_{22} \right\rrbracket - \left\llbracket \mu_{12}^* \right\rrbracket \left\llbracket \mu_{11} \right\rrbracket^{-1} \left\llbracket \mu_{12} \right\rrbracket \right),
    \end{equation}
    \begin{equation}
    \Psi_{24} = \left\llbracket \frac{1}{\sqrt{1 - \kappa^2}} \right\rrbracket^{-1} \cdot
    \left( \left\llbracket \frac{\mu_{23}}{\sqrt{1 - \kappa^2}} \right\rrbracket - \left\llbracket \mu_{12}^* \right\rrbracket \left\llbracket \mu_{11} \right\rrbracket^{-1} \left\llbracket \frac{\mu_{13}}{\sqrt{1 - \kappa^2}} \right\rrbracket \right) ,\; 
    \Psi_{31} = 0,\;  
    \end{equation}
    \begin{equation}
    \Psi_{32} = \left\llbracket \frac{\epsilon_r}{\sqrt{1 - \kappa^2}} \right\rrbracket +\frac{1}{\eta^2} \left\llbracket \sqrt{1 - \kappa^2} \right\rrbracket 
    \llbracket j\pi p \rrbracket \llbracket \mu_{11} \rrbracket^{-1} \llbracket j\pi p \rrbracket ,\; 
    \end{equation}
    \begin{equation}
    \Psi_{33} = \frac{1}{j\eta} \left\llbracket \sqrt{1 - \kappa^2} \right\rrbracket \llbracket j\pi p \rrbracket \llbracket \mu_{11} \rrbracket^{-1} \llbracket \mu_{12} \rrbracket ,\;
    \Psi_{34} = \frac{1}{j\eta} \left\llbracket \sqrt{1 - \kappa^2} \right\rrbracket \llbracket j\pi p \rrbracket \llbracket \mu_{11} \rrbracket^{-1} \left\llbracket \frac{\mu_{13}}{\sqrt{1 - \kappa^2}} \right\rrbracket ,
    \end{equation}
    \begin{equation}
    \Psi_{41} = - \llbracket \epsilon_r \rrbracket \left\llbracket \frac{1}{\sqrt{1 - \kappa^2}} \right\rrbracket^{-1},\;
    \Psi_{42} = 0,\; 
    \Psi_{43} = 0,\; 
    \Psi_{44} = 0,
\end{equation}
\end{widetext}
with

\begin{align}
\left(\left\llbracket \Upsilon \right\rrbracket\right)_{n,m} = \frac{1}{2} \int_{-1}^{1} \Upsilon(\kappa) e^{-j(n-m)\pi \kappa} \, d\kappa
\end{align}
Here, $\left\llbracket \Upsilon \right\rrbracket$ is a square matrix generated by the Fourier coefficients of $\Upsilon$, with $(n,m)$ entries equal to $\Upsilon_{n-m}$; $\eta = k_0 r$.

\section{Expansion of the Incident Field}

The sources of the lowest-order spherical waves are current elements. As it is well-known a single electric-current (an elementary Hertz dipole) with current $I$ and length $l$ parallel to the $z$-axis radiates a $TM_z$ field that can be derived from a  magnetic vector potential. When the dipole is displaced off the origin, a radial wavefunction of the zero-th order is employed and for the radial component of the magnetic vector potential we have
\begin{align}
A_r = r\frac{k_0Il}{j 4\pi h}h_0^{(2)}(k_0|\boldsymbol{r}-\boldsymbol{h}|),
\end{align}
where $\boldsymbol{h}=\boldsymbol{z}h$ and $h_0^{(2)}$ is the spherical Hankel function. Here $h$ is a distance from the global origin to the source location along the $z$-axis. Since (C1) is written in the coordinate system viewed from the source, we need to expand (C1) in spherical wavefunctions referred to the global origin. This can be accomplished using the addition theorem for the spherical functions, and finally we get
\begin{equation}
A_r = r\frac{k_0Il}{j 4\pi h}\sum_{n=0}^{\infty}    (2n+1)h_n^{(2)}(k_0r_>)j_n(k_0r_<)P_n^0(cos\theta)
\end{equation}
with $r_>=max[r,h]$,  $r_<=min[r,h]$ and $j_n$ is the spherical Bessel function of the $n$-th order. The electric and magnetic field components can be expressed through the radial component of the magnetic vector potential \cite{2823} and then can be directly implemented into our formalism (see Sec. III).      

The analysis is straightforward for radially and azimuthally polarized beams. Assuming that the incident beam is propagating along the $z$-axis, we can expand the cylindrical wave functions in terms of spherical ones \cite{stratton2007electromagnetic} 
\begin{equation}
\begin{aligned}
J_m(k_\perp \rho) e^{jk_{||} z}=\sum_{n=0}^{\infty}    j^n(2n+2m+1)
\frac{n!}{(n+2m)!} \cdot \\
P_{n+m}^m(cos\theta) P_{n+m}^m(cos\alpha)j_{n+m}(k_0r).
\end{aligned}
\end{equation}
Here $k_\perp$ and $k_{||}$ are the wavenumbers along the transversal and longitudinal directions with respect to propagation axis, i.e. $z$-axis, $J_m$ is the Bessel function of the first kind with order $n$ and $\alpha$ is an angle between the wavevectors $\boldsymbol{k_0}$ and $\boldsymbol{k_\perp}$. Finally, it is required to expand the $\theta$-dependent terms in (C3) into a 1-D Fourier series as was done in (\ref{eq:7}) and impose the boundary conditions on the spherical shell surfaces to determine the unknown scattering coefficients. No modification of the formalism is needed.

\bibliography{BP}

\end{document}